\documentclass[conference]{IEEEtran}
\usepackage{cases}
\usepackage{bm}
\usepackage[dvips]{graphicx}
\usepackage{subfigure}
\usepackage{amssymb}

\ifCLASSINFOpdf
\else
\fi
\begin{document}
%
\title{Analog Network Coding in the Generalized High-SNR Regime }

\author{\IEEEauthorblockN{Binyue Liu, Ning Cai}
\IEEEauthorblockA{State Key Lab. of ISN, Xidian University\\Xi¡¯an 710071, China\\
Email: {\{liuby, caining}\}@mail.xidian.edu.cn}
}


%


\maketitle

\begin{abstract}
In a recent paper [4], Mari\'c et al. analyzed the performance of the analog network coding (ANC) in a layered relay
network for the high-SNR regime. They have proved that under the ANC scheme, if each relay transmits the received
signals at the upper bound of the power constraint, the transmission rate will approach the network capacity. In this
paper, we consider a more general scenario defined as the generalized high-SNR regime, where the relays at layer $l$ in
a layered relay network with $L$ layers do not satisfy the high-SNR conditions, and then determine an ANC relay scheme
in such network. By relating the received SNR at the nodes with the propagated noise, we derive the rate achievable by
the ANC scheme proposed in this paper. The result shows that the achievable ANC rate approaches the upper bound of the
ANC capacity as the received powers at relays in high SNR increase. A comparison of the two ANC schemes implies that the
scheme proposed in [4] may not always be the optimal one in the generalized
high-SNR regime. The result also demonstrates that the upper and lower bounds of the ANC rate coincide in the limit as
the number of relays at layer L-1 dissatisfying the high-SNR conditions tends to infinity (\textbf{to be infinite}),
yielding an asymptotic capacity result.
\end{abstract}


%
\IEEEpeerreviewmaketitle

\section{Introduction}
Linear network coding [1] achieves the multicast capacity [2] in a noiseless network. This result indicates that each
node
only has to send out a linear combination of its incoming packets. Destination nodes obtain source information
multiplied by a transfer matrix consisting of the global encoding kernels on the incoming edges, and can recover the
original data provided that the matrix is invertible [3].

In a wireless channel, signals simultaneously transmitted from multiple sources add up in the air resulting in
interference. The noisy sum of these signals received at each node may be considered as a linear combination of the
signals and noises. A multihop amplify-and-forward relay scheme referred to as analog network coding extends the concept
of network coding to physical layer [4],[7]. Consequently, in this paper, we define the corresponding local and global
encoding coefficients of ANC. In the high-SNR regime, Mari\'c, Goldsmith, and M\'edard proposed an multihop amplify-and-forward scheme in which
each relay node transmits the received signals at the upper bound of the power constraint. As the main contribution,
they derived the rate achievable by such scheme and showed that it approaches the network capacity in the high-SNR regime.

Gastpar and Vetterli showed a two-hop network model with a joint
source-channel coding relay scheme [5]. The relays
amplify and forward the signals received from the source node with
the amplification gains chosen to achieve the minimum distortion at
the destination. The result shows that the suboptimal scheme in
which the sum transmitting powers of the relays achieves the upper
bound of the sum power constraint is sufficient to approach the
cut-set bound in the limit of large number of relays.

One of the remarkable differences between a multihop
network and a point-to-point channel is that high channel gains do
not necessarily lead to the high-SNR regime [4]. The capacity result is somewhat trivial when all the relay nodes are in high-SNR, whereas it is too complicated to be analyzed when the network has any number of relay nodes with large noises. Hence, we determine the high-SNR condition at each node,
according to which a generalized high-SNR regime for a layered relay network
with $L$ layers is proposed, where the relays at layer $l$ do not
satisfy the high-SNR conditions. Then, the ANC capacity of the layered network is derived in such scenario.
Finally, we analyze a special case when the nodes at the second-to-last layer do
not satisfy the high-SNR condition in the generalized high-SNR
regime. We assume that the transmitting powers of the relays
at that layer are finite constants. With the idea proposed in [5], we also analyze the performance when the number of
nodes tends to infinity. The result shows that the
achievable ANC rate is within a constant gap from the upper bound,
and the gap depends on the number of nodes at that layer.

This paper is organized as follows. A general wireless relay network model and
the definitions of the coding coefficients of ANC are presented in
Section II. The sufficient condition of the power constraint
described by the amplification gains is presented in Section III.
The upper and lower bounds of the ANC rate in the generalized
high-SNR regime are found in Section IV. Two examples demonstrating
the capacity-achieving performance of ANC in the generalized
high-SNR regime are presented in Section V. Section VI concludes the
paper.

The layered network model and most of the definitions
presented in this paper follow those proposed in [4].
The notations used in this paper are as follows. Bold upper- and lower-case
letters denote matrices and column vectors respectively, with
$\left(  \cdot  \right)^T$ denoting their transpose. $E\left[  \cdot  \right]$ is the expectation operation. All the
logarithms in
this paper are in the base 2.

\section{Network Model}
We consider a wireless relay network with a single
source-destination pair shown in Fig.1. We assume that every node k
receives the signal from the source node S through different relay
paths simultaneously. Therefore, all the channel outputs are free of
intersymbol interference. For that reason, we omit the time index in
notations. All nodes are full-duplex. As in [4], the channel output at node k
can be expressed as
\begin{equation}
y_k  = \sum\limits_{j \in N\left( k \right)}
{h_{j,k} x_j  + z_k, }
\end{equation}where $h_{j,k}$ denotes the channel gain from node j to node k, $N\left( k \right)$
represents neighboring nodes of node k, and $z_k$ is the Gaussian
noise with zero mean and variance 1. All the channel gains are
supposed to be fixed real-valued constants and known through the
network for the scope of the present paper.
\begin{figure}
  \centering
  \includegraphics[width=2in,height=1in]{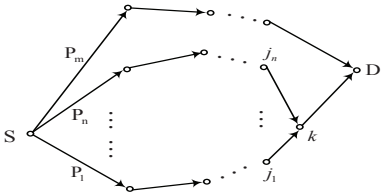}
  \caption{ Wireless relay network.}
\end{figure}
   We assume that there exists a power constraint at node j such that
\begin{equation}
E\left[ {X_j^2 } \right] \le P_j.
\end{equation}

A sequence of codes containing $\left\lceil {2^{nR} } \right\rceil$
codewords of length n is proposed, and it is shown that the error
probability goes to zero as $n \to \infty$
 . In the considered transmission
scheme, the source node encodes with the Gaussian codebook $X_s \sim
\mathcal{N}\left[ {0,P_s } \right]$ , where $\mathcal{N}\left[ {0,\sigma ^2 } \right]$ denotes normal distribution with
zero mean and variance $\sigma^2$. Since each network node performs ANC, node
k transmits:
\begin{equation}
x_k  = \beta _k y_k,
\end{equation}where the amplification gain $\beta _k$ is chosen such that the power constraint (2) is satisfied.
The coding coefficients of ANC are defined as follows.
\newtheorem{definition}{Definition}
\begin{definition}[Local Encoding Coefficient]
Let $e_m$ and $e_{m + 1}$ be the adjacent channels jointed with node k, and
\begin{equation}
\alpha _{e_m ,e_{m + 1} }  = \beta _k h_{e_{m + 1} }
\end{equation}
denotes the local encoding coefficient of pair $\left( {e_m ,e_{m + 1} } \right)$, where, in particular, $
\beta _S  = 1$
  .
\end{definition}

\newtheorem{definition2}{Definition}
\begin{definition}[Global Encoding Coefficient]
 Let
\begin{equation}
f_{j,k}  = \sum\limits_{\left\{ P \right\}} {\prod\limits_{m = 1}^{L\left( P \right)} {\alpha _{e_m ,e_{m + 1} } } }
\end{equation}
 be the global encoding coefficient from node j to node k, where $\{P\}$ represents relay paths between the two nodes,
 and $L\left(P\right)$ denotes the length of $P$.

 The relationship between the local and global coefficients is as follows.
\begin{equation}
f_{S,k}  = \sum\limits_{j \in N\left( k \right)} {\beta _j h_{j,k} f_{S,j} }
\end{equation}
\end{definition}

\section{Sufficient Condition of Power Constraint}
In a wireless relay network, each network node can be considered as a source of noise. The network can be illustrated as
a combination of several equivalent single source-destination graphs. Therefore, the signal received at node k is
expressed as
\begin{equation}
y_k  = f_{{\rm{S}},k} x_{\rm{S}}  + \sum\limits_{\left\{ {i_k } \right\}} {f_{i_k ,k} z_{i_k } }  + z_k,
\end{equation}
where $\{i_k\}$ denotes nodes in the relay paths from the source node S to node k.
\newtheorem{definition3}{Definition}
\begin{definition}
When each node j transmits with $P_j$ given in (2), the power received at node k is denoted as
\begin{equation}
P_{R,k}  = \left( {\sum\limits_{j \in N\left( k \right)} {h_{j,k} \sqrt {P_j } } } \right)^2,
\end{equation}
and the reciprocal of $P_{R,k}$ is represented by
\begin{equation}
\delta _k  = \frac{1}{{P_{R,k} }}.
\end{equation}
\end{definition}

We extend the sufficient condition of the power constraint in [4] to the general scenario described in section II by the
following theorem.

\newtheorem{thm}{Theorem}
\begin{thm}
At every node performing ANC with the amplification gain
\begin{equation}
\beta _k^2  \le \frac{{P_k }}{{\left( {1 + \delta _k } \right)P_{R,k} }}
\end{equation}
the power constraint (2) is satisfied.
\end{thm}

\begin{IEEEproof}
Let $w_k$ be the total noise received at node k.
\begin{equation}
w_k  = \sum\limits_{\left\{ {i_k } \right\}} {f_{i_k ,k} z_{i_k } }  + z_k
\end{equation}
We prove the theorem by induction. Consider first the node k whose neighboring nodes set only contains the source node
S. From (1) and (3),
\begin{equation}
E\left[ {X_k^2 } \right] = \beta _k^2 E\left[ {Y_k^2 } \right]\mathop  \le \limits^{\left( a \right)} \frac{{P_k
}}{{\left( {1 + \delta _k } \right)P_{R,k} }}\left( {h_{S,k}^2 P_{\rm{S}}  + 1} \right)\mathop  = \limits^{\left( b
\right)} P_k,
\end{equation}
where (a) is obtained from the condition (10), and (b) is drawn from (8).
To prove that the theorem holds for any node, we assume that all the neighboring nodes of node k satisfy the power constraints such that
\begin{equation}
E\left[ {X_j^2 } \right] = E\left[ {\beta _j^2 \left( {f_{S,j} x_S  + w_j } \right)^2 } \right] \le P_j ,j \in N\left( k
\right)
\end{equation}
Then we consider the transmitting power at node k,
\begin{equation}
E\left[ {X_k^2 } \right]\mathop  = \limits^{\left( a \right)} E\left[ {\beta _k^2 \left( {f_{S,k} x_S  + \sum\limits_{j
\in N\left( k \right)} {\beta _j h_{j,k} w_j }  + z_k } \right)^2 } \right]
\end{equation}

\setlength{\arraycolsep}{0.0em}
\begin{eqnarray}
\mathop  \le \limits^{\left( b \right)} &&c_{00}\left\{ {f_{S,k}^2 P_S  + E\left[ {\left( {\sum\limits_{j \in N\left( k
\right)} {\beta _j h_{j,k} w_j } } \right)^2 } \right]+1} \right\}\\
\mathop  = \limits^{\left( c \right)}&&c_{00}\left\{ {c_{01}  + \sum\limits_{j \in N\left( k \right)} {h_{j,k}^2 \beta
_j^2 \left[ {f_{S,j}^2 P_S  + E\left( {w_j^2 } \right)} \right]} } \right\} + c_{00}\\
\mathop  \le \limits^{\left( d \right)} &&c_{00}\left\{ {c_{01}  + \sum\limits_{j \in N\left( k \right)} {h_{j,k}^2 P_j
} } \right\} + c_{00}\\
\mathop  \le \limits^{\left( e \right)} &&c_{00}\left\{ {\sum\limits_{l,j \in N\left( k \right),l \ne j} {\beta _j
h_{j,k} \beta _l h_{l,k} } } \right.\left[ {\sqrt {E\left( {w_j^2 } \right)E\left( {w_l^2 } \right)} }
\right.\nonumber\\
&&\left. { + f_{S,j} f_{S,l} P_S } \right]\left. { + \sum\limits_{j \in N\left( k \right)} {h_{j,k}^2 P_j } } \right\} +
c_{00}\\
\mathop  \le \limits^{\left( f \right)}&& c_{00}\left\{ {\sum\limits_{j \in N\left( k \right)} {h_{j,k}^2 P_j }  +
\sum\limits_{l,j \in N\left( k \right),l \ne j} {h_{j,k} h_{l,k} } } \right.\left. { \sqrt {P_j P_l } } \right\} +
c_{00}\nonumber\\
\\
=&&P_k
\end{eqnarray}where $c_{00}  = \displaystyle \frac{{P_k }}{{\left( {1 + \delta _k } \right)P_{R,k} }}$,
\[
c_{01}=\sum\limits_{l,j \in N\left( k \right),l \ne j} {\beta _j h_{j,k} \beta _l h_{l,k} \left[ {E\left( {w_j w_l }
\right) + f_{S,j} f_{S,l} P_S } \right]},
\]and

(a)	follows from (6), (7)and (11),

(b)	follows from (10),

(c)	follows from (6),

(d)	follows from the assumption (13),

(e)	follows from the Schwarz inequality,

(f)  also follows from (13), and the fact that\\*$\left[ {\sqrt {E\left( {w_j^2 } \right)E\left( {w_l^2 } \right)} +
f_{S,j} f_{S,l} P_S} \right]^2$\[\le \left[ {E\left( {w_j^2 } \right) + f_{S,j}^2 P_S } \right]\left[ {E\left( {w_l^2 }
\right) + f_{S,l}^2 P_S } \right].\]
Then we complete the proof.
\end{IEEEproof}

\section{Upper and Lower Bounds to ANC Rate}

\subsection{Layered Network in Generalized High-SNR Regime}
\begin{figure}
  \centering
  \includegraphics[width=2in,height=1.2in]{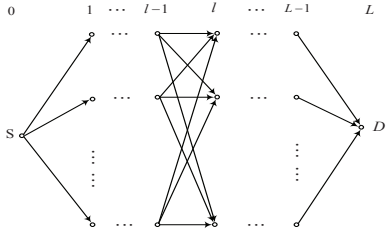}
  \caption{ L-layer relay network.}
\end{figure}
We consider a layered relay network with the source node at layer 0 and the destination node at layer L shown in Fig. 2. As in [4], the received signal vector at layer $l+1$ is
 \begin{equation}
{\bf{y}}_{l + 1}  = {\bf{H}}_l {\bf{x}}_l {\bf{ + z}}_{l + 1}.
\end{equation}
The goal is to find suitable $
\beta _j
$'s to maximize the SNR at the destination. The optimization problem is defined as follows and the SNR function is obtained from (7).
\begin{equation}
\left\{ {\begin{array}{*{20}c}
   {\max {\rm{ }}SNR\left( {\left\{ {\beta _j ,j \in 1,2 \cdots L - 1} \right\}} \right)
 = \frac{{{f_{S,D}^2} P_S }}{{\sum\limits_{\left\{ {i_D } \right\}} {f_{i_D ,D}^2 }  + 1}}}  \\
   {s.t.\beta _j^2  \le \frac{{P_j }}{{\left( {1 + \delta _j } \right)P_{R,j} }},j \in 1,2 \cdots L - 1}  \\
\end{array}} \right.{\rm{ }}.
\end{equation}Maximizing the expression (22) over all amplification gains under the given power constraints does not
seem to have a simple solution. However, a good solution is found after we further limit the network model in the
generalized high-SNR regime defined as follows.

\newtheorem{definition4}{Definition}
\begin{definition}
 The high-SNR condition at node j is defined as
 \begin{equation}
P_{R,j}  \ge \frac{1}{\delta }
\end{equation}
for some small $
\delta  \ge 0
$.
\end{definition}

\newtheorem{definition5}{Definition}
\begin{definition}
 The wireless network is in the generalized high-SNR regime if
 \begin{equation}
\mathop {\min }\limits_{j \in 1, \cdots ,l - 1,l + 1, \cdots ,L} P_{R,j}  \ge \frac{1}{\delta },
\end{equation}
\end{definition}
which implies except the nodes in layer $l(l = 1,2, \cdots ,L)$, any other nodes are in the high-SNR condition.
\newtheorem{remark}{Remark}
\begin{remark}
For the purpose of this paper, we assume that the received SNR at each node $
k,k \notin l
$
is large. The general scenario defined here can be considered as an extension of the network model discussed in [4]. Here, if we set $
l = L
$, the generalized high-SNR regime is the high-SNR regime.
\end{remark}

\subsection{Upper Bound to ANC Capacity}
An upper bound to the ANC capacity of the network shown in Fig. 2 in the generalized high-SNR regime can be found from
an analysis of an ideal network, where except the $
lth
$
layer (here we assume that $
l \ne L
$), any other layers are noiseless. We draw the conclusion in the following theorem.

\newtheorem{thm2}{Theorem}
\begin{thm}
The upper bound of the ANC capacity in the generalized high-SNR regime (24) of the Gaussian relay network with the
transmitting power constraints (2) is
\begin{equation}
R_1  = \frac{1}{2}\log \left( {1 + {\bf{P}}_{R,l}^T {\bf{P}}_{R,l} } \right),
\end{equation}
where ${\bf{P}}_{R,l}  = \left[ {\sqrt {P_{R,1} }  \cdots \sqrt {P_{R,j} }  \cdots \sqrt {P_{R,n_l } } } \right]^T ,j
\in l.$
\end{thm}
\begin{IEEEproof}
We denote the SNR function at the destination of this ideal network as $SNR_1$. With the identical amplification gains,
we have
\begin{equation}
SNR_1 \left( {\left\{ {\beta _j ,j \in 1 \cdots L - 1} \right\}} \right) \ge SNR\left( {\left\{ {\beta _j ,j \in 1
\cdots L - 1} \right\}} \right)
\end{equation}
Clearly, the maximum of $SNR_1$ is no less than the maximum of $SNR$.
We first point out that
\begin{equation}
{\bm{\gamma }} = {\bf{B}}_l {\bf{g}} = {\bm{G\beta }}_l,
\end{equation}
where ${\bm{\beta }}_l  = \left[ {\beta _1  \cdots \beta _j  \cdots \beta _{n_l } } \right]^T ,j \in l$, ${\bf{B}}_l  =
{\rm{diag}}\left\{ {\beta {}_1 \cdots \beta {}_j \cdots \beta {}_{n_l }} \right\},j \in l$, ${\bf{g}} = \left(
{{\bf{h}}_{L - 1}^T {\bf{B}}_{L - 1}  \cdots {\bf{H}}_l } \right)^T  = \left[ {g_1  \cdots g_j  \cdots g_{n_l } }
\right]^T$, and ${\bf{G}} = {\rm{diag}}\left\{ {g_1  \cdots g_j  \cdots g_{n_l } } \right\}
$. Then we denote the correlation matrix of the signal vector received at layer $l$ by
\begin{equation}
E\left[ {{\bf{y}}_l {\bf{y}}_l^T } \right] = {\bf{P}}_l {\bf{P}}_l^T.
\end{equation}
Therefore, $SNR_1$ can be expressed as
\begin{equation}
SNR_1 \left( {\bm{\gamma }} \right) = \frac{{{\bm{\gamma }}^T {\bf{P}}_l {\bf{P}}_l^T {\bm{\gamma }}}}{{{\bm{\gamma }}^T
{\bm{\gamma }}}}.
\end{equation}
With the power constraint (2), the sum signal power received at layer $l$ is upper bounded by $
{\bf{P}}_{R,l}^T {\bf{P}}_{R,l}
$. The maximum of $SNR_1$ is
\begin{equation}
\max SNR_1 \left( {{\bm{\gamma }}_{opt} } \right) = {\bf{P}}_{R,l}^T {\bf{P}}_{R,l},
\end{equation}and the Gaussian channel capacity evaluates to
\[
R_1  = \frac{1}{2}\log \left( {1 + {\bf{P}}_{R,l}^T {\bf{P}}_{R,l} } \right).
\]
Then we complete the proof.
\end{IEEEproof}

We propose an ANC scheme such that the amplification gain at node $
j,j \in 1,2 \cdots l - 1{\rm{,}}l + 1 \cdots L - 1
$ is chosen as
\begin{equation}
\beta _j^2  = \frac{{P_j }}{{\left( {1 + \delta } \right)P_{R,j} }},j \in 1,2 \cdots l - 1{\rm{,}}l + 1 \cdots L - 1,
\end{equation}
and at node $
k,k \in l
$
is chosen as
\begin{equation}
{\bm{\beta }}_l  = c_1 {\bf{G}}^{ - 1} {\bf{P}}_{R,l},
\end{equation}
where $c_1$ can be determined as
\begin{equation}
c_1  = \min \left\{ {\frac{{g_j }}{{P_{R,j} }}\sqrt {\frac{{P_j }}{{1 + \delta _j }}} ,j = 1,2 \cdots n_l } \right\}.
\end{equation}
Using this scheme in (29), the lower bound of achievable ANC rate in such ideal network results.
\begin{equation}
R = \frac{1}{2}\log \left( {1 + \frac{1}{{\left( {1 + \delta } \right)^{l - 1} }}{\bf{P}}_{R,l}^T {\bf{P}}_{R,l} }
\right).
\end{equation}
As $
\delta  \to 0
$, the achievable ANC rate approaches the upper bound of the ANC capacity (25).

\newtheorem{remark2}{Remark}
\begin{remark}
Evidently, the case when $
l = L
$, the scheme proposed in [4] can be considered as a special case of the previous general scenario. Since the noise power received at the destination is independent of amplification gains, the larger the transmitting powers, the better the performance of the ANC scheme. With the power constraint, the optimal amplification gain at each node is
\begin{equation}
\beta _j  = \sqrt {\frac{{P_j }}{{\left( {1 + \delta _j } \right)P_{R,j} }}}
\end{equation}
The corresponding lower bound of the ANC rate is
\begin{equation}
R = \frac{1}{2}\log \left( {1 + \frac{1}{{\left( {1 + \delta } \right)^{L - 1} }}P_{R,D} } \right)
\end{equation}
which approaches the MAC cut-set bound $
C = \frac{1}{2}\log \left( {1 + P_{R,D} } \right)
$ as $
\delta  \to 0
$.
\end{remark}
\newtheorem{remark3}{Remark}
\begin{remark}
By the assumption, we observe that the bottleneck on the data transfer is on the cut-set between the $
\left( {l - 1} \right)th
$ and the $lth$ layers. We assume the nodes at layer $l$ and layer $l+1$ are the multiple transmitting antennae of the
source node and the multiple receiving antennae of the destination node respectively. In particular, when $
rank\left( {{\bf{H}}_{l - 1}^T {\bf{H}}_{l - 1} } \right) = 1
$, the MIMO capacity which can be seen as the cut-set bound [6],[8] evaluates to
\begin{equation}
C = \frac{1}{2}\log \left( {1 + {\bf{P}}_{R,l}^T {\bf{P}}_{R,l} } \right),
\end{equation}
which equals the upper bound of the ANC capacity. However, the hypothesis that $
rank\left( {{\bf{H}}_{l - 1}^T {\bf{H}}_{l - 1} } \right) = 1
$ does not always hold in general scenario. Hence, the cut-set bound should not be expected to be tight.
\end{remark}
\subsection{Lower Bound to ANC Rate Achievable}
Lower bound to ANC rate achievable in the generalized high-SNR regime is found by analyzing the performance of the ANC
scheme described in the previous section. The lower bound of ANC rate is derived in the following theorem.
\newtheorem{thm3}{Theorem}
\begin{thm}
The achievable rate of the Gaussian layered relay network of Fig.2 in the generalized high-SNR regime by the ANC scheme under the transmitting power constraints is lower bounded by
\begin{equation}
R_2  = \frac{1}{2}\log \left( {1 + \frac{{\frac{{{\bf{P}}_{R,l}^T {\bf{P}}_{R,l} }}{{\left( {1 + \delta } \right)^{l - 1} }}}}{{\left[ {1 - \frac{1}{{\left( {1 + \delta } \right)^{l - 1} }}} \right]{\bf{P}}_{R,l}^T {\bf{P}}_{R,l}  + c_3 }}} \right),
\end{equation}where $
c_3  = 1 + \displaystyle\frac{{c_2 }}{{c_1 ^2 {\bf{P}}_{R,l}^T {\bf{P}}_{R,l} }}
$ and
$
c_2  = \sum\limits_{d = 1}^{L - \left( {l + 1} \right)} {\displaystyle\frac{{\delta P_{R,D} }}{{\left( {1 + \delta }
\right)^d }}}  + 1$.
\end{thm}
\begin{remark}
With the previous ANC scheme, the SNR function at the destination can be calculated. Assuming that the noises at the same layer, except the ones at layer $l$, are dependent, the sum noise power at the destination is enlarged. Then the result is derived. The details of the proof are omitted.
\end{remark}
%
%

The case when the received powers of nodes in high-SNR condition tend to infinity at the same rate, i.e., $\delta  \to 0$, and $P_{R,k}  = {\rm{const}}.,k \in l$ as $\delta  \to 0$.
The limit of the difference between the two bounds is
\begin{equation}
 \mathop {\lim }\limits_{\delta  \to 0} (R_1-R_2)  = 0.
\end{equation}
The result implies that the achievable ANC rate approaches the ANC capacity (25).

Finally, we discuss another special case when $l=L-1$. Suppose that the transmitting powers of the nodes at this layer
are finite constants. The received signal power at the destination increases with the number of the nodes at this layer.
Let $
\delta ' =  \frac{1}{{\mathop {\min }\limits_{j \in 1,2 \cdots L - 2} P_{R,j} }}
$ be some small positive value, which implies the nodes in those layers satisfy the high-SNR conditions. With
sufficiently large number of the nodes at layer $L-1$, the received signal power at the destination also satisfies the
high-SNR condition. Therefore, the network model can be considered in the generalized high-SNR regime.

$\mathop {\lim }\limits_{n_l  \to \infty } \mathop {\lim }\limits_{\delta ' \to 0} (R_1-R_2)$
\begin{equation}
= \mathop {\lim }\limits_{n_l  \to \infty } \displaystyle \frac{1}{2}\log \left( {\frac{{1 + {\bf{P}}_{R,L - 1}^T
{\bf{P}}_{R,L - 1} }}{{1 + \displaystyle \frac{{{\bf{P}}_{R,L - 1}^T {\bf{P}}_{R,L - 1} }}{{1 + \displaystyle
\frac{1}{{c_1 ^2 {\bf{P}}_{R,L - 1}^T {\bf{P}}_{R,L - 1} }}}}}}} \right) = 0.
\end{equation}

The result implies that the ANC rate behaves asymptotically like $
C = \frac{1}{2}\log \left( {1 + {\bf{P}}_{R,L-1}^T {\bf{P}}_{R,L-1} } \right)
$.

\section{Examples}

\begin{figure}
  \centering
  \includegraphics[width=1.5in,height=1in]{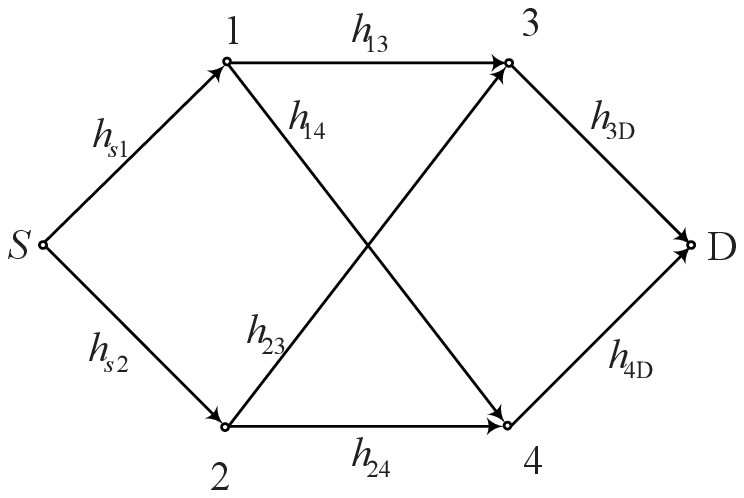}
  \caption{3-layer network with 2 nodes at layer 2.}
\end{figure}
\begin{figure}
  \centering
  \includegraphics[width=2.4in]{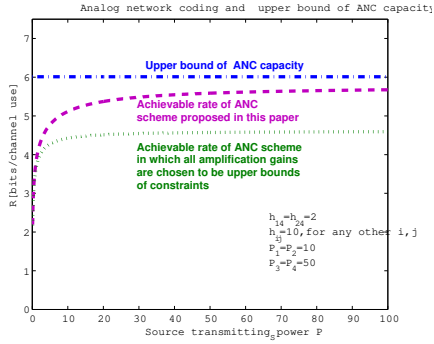}
  \caption{ Achievable ANC rate and upper bound of ANC capacity in 3-layer network.}
\end{figure}
\noindent\emph{Example 1: 3-Layer Network with 2 relays at layer 2}

For the network shown in Fig.3, we present the performance of ANC in a 3-layer network in the generalized high-SNR
regime with nodes at layer 2 dissatisfying the high-SNR conditions.
From Fig.4 we observe the following:

1)	For $P_{R,3}$ and $P_{R,4}$ constants, the achievable ANC rate approaches the upper bound of ANC capacity. The ANC
rate approaches the capacity to within one bit as $P_S  > 10$, and is within a small fraction of a bit for $P_S  >
100$.

2)	Fig.4 also shows the achievable ANC rate by setting the amplification gains to the upper bounds of the power
constraints. There is a constant gap approximate 1.5 bits from the ANC capacity.

\noindent\emph{Example 2: 3-Layer Network with n relays at layer 2}

We next present the performance of ANC in the 3-layer network with n relays at layer 2 shown in Fig.5.

The result shown in Fig. 6 implies that the gap between the achievable ANC rate and the upper bound of the ANC capacity
decreases within one bit as $n > 5$, and within a small fraction of a bit for $n > 40$ as the transmitting powers of the
nodes at layer 2 are limited to 2.
\begin{figure}
  \centering
  \includegraphics[width=1.7in,height=1in]{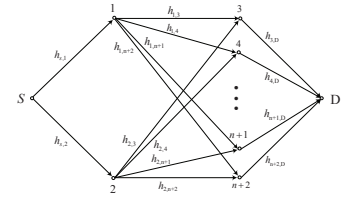}
  \caption{3-layer network with n nodes at layer 2.}
\end{figure}
\begin{figure}
  \centering
  \includegraphics[width=2.4in]{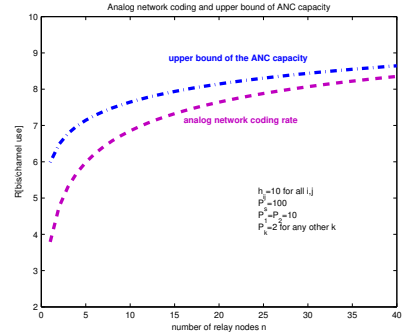}
  \caption{Upper bound of ANC capacity and increase of achievable ANC rate with number of relays at layer 2 in 3-layer
  network.}
\end{figure}

\section{conclusion}
We derived the capacity of a multihop network by the ANC scheme in the generalized high-SNR regime. As all the received powers of the nodes in the high SNR increase, the achievable ANC rate by the ANC scheme proposed in this paper approaches the upper bound of the ANC capacity. As discussed in this paper, we assume that all the channel gains are fixed real-valued constants and known through the network. Relaxing this assumption to a fading scenario is a topic for our future work.


\section*{Acknowledgment}
The authors are supported by grants from the National
Natural Science Foundation of China (60832001).




\begin{thebibliography}{1}

\bibitem{IEEEhowto:kopka}
S.-Y. R. Li, R. W. Yeung, and N. Cai, "Linear network coding," IEEE Trans. Inf. Theory, vol. 49, no. 2, pp. 371-381,
Feb. 2003.
\bibitem{IEEEhowto:kopka}
R. Ahlswede, S. L. N. Cai, and R. Yeung, "Network information flow," IEEE Trans. Inf. Theory, vol. 46, no. 4, pp.
1204-1216, Jul.2000.
\bibitem{IEEEhowto:kopka}
R. Koetter and M. M\'edard, "An algebraic approach to network coding," IEEE/ACM Trans. Networking, vol. 11, no. 5, pp.
782-795,Oct. 2003
\bibitem{IEEEhowto:kopka}
I. Mari\'c, A. Goldsmith, and M. M\'edard, "Analog network coding in the high-SNR regime," in Wireless Network Coding
Workshop, pp. 1 - 6, Jun. 2010.
\bibitem{IEEEhowto:kopka}
M. Gastpar and M. Vetterli, "On the capacity of large Gaussian relay networks," IEEE Trans. Inf. Theory, vol. 51, no. 3,
pp. 765-779, Mar. 2005.
\bibitem{IEEEhowto:kopka}
\.I. E. Telatar, "Capacity of multi-antenna Gaussian channels," Bell Labs, Tech. Memo., 1995.
\bibitem{IEEEhowto:kopka}
S. Katti, S. Gollakota, and D. Katabi, "Embracing wireless interference: Analog network coding," in ACM SIGCOMM, 2007.
\bibitem{IEEEhowto:kopka}
T. Cover and J. Thomas, Elements of Information Theory. John Wiley Sons, Inc., 1991.

\end{thebibliography}
%

\end{document}